\documentstyle[twocolumn,aps,epsf]{revtex}
\begin{document}
\newcommand{\beq}{\begin{equation}}
\newcommand{\eeq}{\end{equation}}

\title{CHIRAL GLASS PHASE IN CERAMIC SUPERCONDUCTORS}

\author{Mai Suan Li$^1$ and Hikaru Kawamura$^2$}

\address{$^1$Institute of Physics, Polish Academy of Sciences,
Al. Lotnikow 32/46, 02-668 Warsaw, Poland\\
$^2$ Faculty of Engineering and Design, Kyoto Institute
of Technology, Sakyo-ku, Kyoto 606 }

\address{
\centering{
\medskip\em
{}~\\
\begin{minipage}{14cm}
A three-dimensional lattice
of the Josephson junctions with a finite self-conductance is employed
to model the ceramic superconductors.
By Monte Carlo simulations it is shown that the chiral glass
phase is stable in three dimensions even under the influence of
screening.
The nonlinear ac susceptibility and the compensation effect 
are also studied.
The compensation effect is shown to be due to the existence of 
the chiral glass phase.
In agreement with experiments, 
this effect is demonstrated to be
present in the ceramic superconductors which show the paramagnetic 
Meissner effect.
{}~\\
{}~
\end{minipage}
}}

\maketitle



\section{Introduction}

Since the discovery ceramic high-$T_c$ superconductors 
it has been known
that they may exhibit a glassy behavior
reminiscent of the spin glass \cite{Muller}.
Recently, by the noise and ac susceptibility measurements Leylekian
{\em et. al.} demonstrated that LSCO ceramics show the glassy behavior 
{\em even in zero external field} \cite{Leylekian}.
They also observed an intergranular cooperative phenomenon indicative
of a glassy phase transition. This collective phenomenon
may be interpreted in terms of the {\em chiral glass} (CG) picture
\cite{Kawa1} using 
a three-dimensional lattice model
of the Josephson junctions with a finite self-conductance. 
The order parameter of the CG phase is a "chirality" which 
represents the direction of the local loo-supercurrent over grains
\cite{Dominguez,KawLi}. The frustration essential to realize the CG phase
arises due to random distribution of 0- and $\pi$-junctions with positive
and negative Josephson couplings, respectively \cite{Sigrist}.
In this paper we review our recent results obtained by
Monte Carlo simulations on the nature of ordering
of the CG. One of the most important conclusions is that the screening
effect does not destroy this phase in three dimensions
(the vortex glass phase which
may exist only in the
non-zero external field \cite{Fisher} is, in contrast,
unstable under influence
of the screening \cite{Bokil}). It should be noted that more direct
support of the CG has been reported by the ac susceptibility \cite{Matsuura}
and resistivity \cite{Matsuura1} measurements on 
YBCO ceramic samples and by
the aging effect in Bi$_2$Sr$_2$CaCu$_2$O$_8$ \cite{Nordblat}.

Another issue of the present paper is to explain the so called compensation
effect (CE) observed in  some ceramic superconductors \cite{Heinzel}.
Overall, this effect  may be detected in the following way.
The sample is cooled in
the external dc field down to a low temperature and then the field is
switched off. At the fixed low $T$ the second harmonics are monitored
by applying the dc and ac
fields to the sample.
Due to the presence of non-zero spontaneous orbital
moments the remanent magnetization or, equivalently, the internal field appears
in the cooling process.
If the direction of the external dc field is
identical to that during the field cooled (FC) procedure, the induced
shielding
currents will reduce the remanence. Consequently, the absolute value
of the second harmonics $|\chi_2|$ decreases until
the signal of the second harmonics is minimized
at a field $H_{dc}=H_{com}$.
Thus the CE is a phenomenon in which the external and internal fields are
compensated and the second harmonics become zero.

The key observation of Heinzel {\em et al.}\cite{Heinzel}
is that the CE
appears only in the
samples which show the paramagnetic Meissner effect (PME)
\cite{Sved,Braunish} but not in those which do not.
It should be noted that
the intrinsic mechanism leading to the PME is still under
debate\cite{Geim,Sigrist1}.
Sigrist and Rice argued that the PME in the high-$T_c$ 
superconductors is consistent with the $d$-wave superconductivity\cite{Sigrist}.
On the other hand, the paramagnetic response
has been seen even in the conventional Nb \cite{Thompson,Kostic,Pust} 
and Al \cite{Geim} superconductors.
In order to explain the PME in terms of conventional superconductivity
one can employ the idea of the flux compression inside of a sample. Such
phenomenon becomes possible in the presence of the inhomogeneities\cite{Larkin}
or of the sample boundary\cite{Moshchalkov}.

In this paper we explain the CE theoretically by 
Monte Carlo simulations.
Our starting point is based on the possible existence of the CG
phase in which the remanence necessary for observing the CE 
should occur in the cooling procedure.
Such remanence phenomenon is similar to what happens in spin glass.
In fact, in the CG phase
the frustration due to existence of 0- and $\pi$-junctions
leads to non-zero supercurrents. 
The internal field (or the remanent 
magnetization) induced by the supercurrents
in the cooling process from high temperatures to
the CG phase may compensate the external dc field.

Using the three-dimensional
XY model of the Josephson network 
with  finite self-inductance
We show that in the FC regime the CE appears 
in the samples 
which show the PME.
This finding
agrees with the experimental data of Heinzel {\em et al}\cite{Heinzel}.

\section{Model}

We neglect the charging effects of the grain and
consider the following Hamiltonian\cite{Dominguez,KawLi}
\begin{eqnarray}
{\cal H} = - \sum _{<ij>} J_{ij}\cos (\theta _i-\theta _j-A_{ij})+ \nonumber\\
\frac {1}{2{\cal L}} \sum _p (\Phi_p - \Phi_p^{ext})^2, \nonumber\\
\Phi_p \; \; = \; \; \frac{\phi_0}{2\pi} \sum_{<ij>}^{p} A_{ij} \; , \;
A_{ij} \; = \; \frac{2\pi}{\phi_0} \int_{i}^{j} \, \vec{A}(\vec{r}) 
d\vec{r} \; \; ,
\end{eqnarray}
where $\theta _i$ is the phase of the condensate of the grain
at the $i$-th site of a simple cubic lattice,
$\vec A$ is the fluctuating gauge potential at each link
of the lattice,
$\phi _0$ denotes the flux quantum, 
$J_{ij}$ denotes the Josephson coupling
between the $i$-th and $j$-th grains, 
${\cal L}$ is the self-inductance of a loop (an elementary plaquette),
while the mutual inductance between different loops
is neglected.
The first sum is taken over all nearest-neighbor pairs and the
second sum is taken over all elementary plaquettes on the lattice.
Fluctuating  variables to be summed over are the phase variables,
$\theta _i$, at each site and the gauge variables, $A_{ij}$, at each
link. $\Phi_p$ is the total magnetic flux threading through the 
$p$-th plaquette, whereas $\Phi_p^{ext}$ is the flux due to an 
external magnetic field applied along the $z$-direction,
\begin{equation}
\Phi_p^{ext} = \left\{ \begin{array}{ll}
                   HS \; \;  & \mbox{if $p$ is on the $<xy>$ plane}\\
                   0  & \mbox{otherwise} \; \; ,
                        \end{array}
                  \right. 
\end{equation}
where $S$ denotes the area of an elementary plaquette.
In what follows we assume $J_{ij}$ to be an independent random variable
taking the values $J$ or $-J$ with equal probability ($\pm J$ or bimodal
distribution), each representing 0 and $\pi$ junctions.

\section{Existence of CG phase}

In this section we employ the finite size scaling technique
to study the nature of ordering of the CG phase in three dimension.
The external field is set to be equal to zero ($\Phi_p^{ext}$=0).

At each plaquette the local chirality is defined by the gauge-invariant
quality,
\begin{equation}
\kappa \; = \; 2^{-3/2} \sum_{<ij>}^p \tilde{J}_{ij}
\sin (\theta _i-\theta _j-A_{ij}),
\end{equation}
where $\tilde{J}_{ij}=J_{ij}/J$ and the sum runs over a directed contour
along the sides of the plaquette $p$.
The overlap between two replicas of the chirality is 
\begin{equation}
q_{\kappa} \; = \; \frac{1}{N_p} \sum_{p} \kappa _p^{(1)}
\kappa _p^{(2)},
\end{equation}
where $N_p$ is the total number of plaquettes.
In term of this chiral overlap, the Binder ratio of the chirality is defined
as follows
\begin{equation}
g_{CG} \; = \; \frac{1}{2} ( 3 - 
[<q_{\kappa }^4>]/[<q_{\kappa }^2>]^2 ).
\end{equation}
Here $<...>$ and $[...]$ represent the thermal and the configuration
average, respectively. $g_{CG}$ is normalized so that it tends to zero above 
the chiral-glass transition temperature, $T_{CG}$, and tends to unity below
$T_{CG}$ provided the ground state is non-degenerate.

\begin{figure}
\epsfxsize=3.2in
\centerline{\epsffile{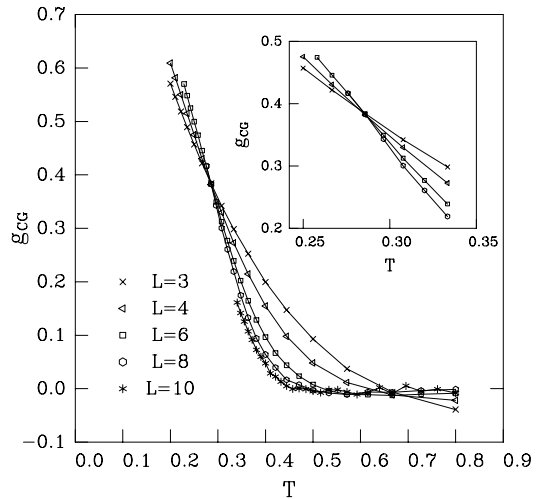}}
\caption{The temperature and size dependence of the Binder ratio
$g_{CG}$ for
$\tilde{\cal{L}}=1$. Inset is a magnified view around the transition 
temperature $T_{CG}=0.286 \pm 0.01$. }
\end{figure}

Monte Carlo simulation is performed according to the replica exchange
method \cite{Hukushima}. Equilibration is checked by monitoring
the stability of the results against at least three-times longer runs 
for a subset of samples. The free boundary conditions are employed.
Fig. 1 shows the results for the system sizes
$L=3, 4, 6, 8$ and 10, and for the dimensionless inductance
$\tilde{\cal{L}}=1$ ($\tilde{\cal{L}}=(2\pi /\phi _0)^2J\cal{L}$).
The number of samples we used is running from 1500 to 100
depending on the system size $L$.
Obviously, all of the curves of $g_{CG}$ for $L=3, 4, 6$ and 8 
cross at almost the same temperature, strongly suggesting the occurence
of a finite-temperature CG transition at $T_{CG}=0.286\pm 0.01$.
A more careful analysis \cite{Kawa1} shows that the CG phase exists
if $\tilde{\cal{L}}$ is smaller than a critical value $\tilde{\cal{L}}_c$,
where $5 \leq \tilde{\cal{L}}_c \leq 7$.
Thus, the CG phase is stable against the screening if the latter 
is not strong enough.

\section{Compensation effect}

In order to study the CE one has to apply the external field $H$ 
which includes the dc and ac parts
\begin{equation}
H \; \; = \; \; H_{dc} + H_{ac} \cos(\omega t) \; \; .
\end{equation}
It should be noted that the dc field is necessary to generate even harmonics.
The ac linear susceptibilty of model (1)
has been studied\cite{KawLi} by Monte Carlo simulations. 
Here we go beyond our previous calculations of the linear ac susceptibility
\cite{KawLi}. We study the dependence of the second harmonics
as a function of the dc field. In this way, we can make a direct comparison
with the CE observed in the experiments \cite{Heinzel}.

The dimensionless magnetization along the $z$-axis mormalized per plaquette,
$\tilde{m}$, is given by
\begin{equation}
\tilde{m} \; \; = \; \; \frac{1}{N_p\phi_0} \; \sum_{p\in <xy>} 
(\Phi_p - \Phi_p^{ext}) \; \; ,
\end{equation}
where the sum is taken over all $N_p$ plaquettes on the $<xy>$ plane of the
lattice. The real and imaginary parts of the ac second order susceptibility
$\chi'_2(\omega)$ and $\chi''_2(\omega)$ are calculated as
\begin{eqnarray}
\chi'_2(\omega) \; \; &=& \; \; \frac{1}{\pi h_{ac}}
\int_{-\pi}^{\pi} \; \tilde{m}(t) \cos(2\omega t)d(\omega t) \; \; , \nonumber\\
\chi''_2(\omega) \; \; &=& \; \; \frac{1}{\pi h_{ac}}
\int_{-\pi}^{\pi} \; \tilde{m}(t) \sin(2\omega t)d(\omega t) \; \; ,
\end{eqnarray}
where $t$ denotes the Monte Carlo time. The dimensionless ac field 
$h_{ac}$, dc field $h_{dc}$  
are defined 
as follows 
\begin{eqnarray}
h_{ac} \; \; = \; \; \frac{2\pi H_{ac}S}{\phi_0} \; \; , \; \;
h_{dc} \; \; = \; \; \frac{2\pi H_{dc}S}{\phi_0} \; \; , \; \;
\end{eqnarray}

\begin{figure}
\epsfxsize=3.2in
\centerline{\epsffile{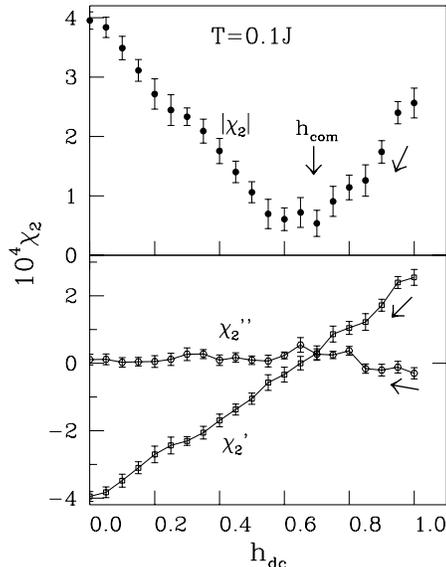}}
\caption{
The second harmonics obtained
after field cooling in a dc field $h_{dc}=1$ from $T=0.7$ to $T=0.1$.
The temperature is reduced in steps of 0.05.
At the lowest $T=0.1$ the dc field used in cooling is switched off
and the second harmonics are generated by applying the combined
field (6). The $dc$ field is stepwise reduced from $h_dc=1$ to $h_dc=0$.
The inductance is chosen to be equal to $\tilde{{\cal L}}=4$.
The arrows indicate the sense of the changes in the
dc field. The results are averaged over 40 samples and are
qualitatively the same as those presented in Fig. 1 of Ref. [12].}
\end{figure}

For model (1) the PME appears clearly 
for $h_{dc} \leq 1$ \cite{KawLi}. So the largest $h_{dc}$
we take is 1. On the other hand, as mentioned above,
the CG phase 
is found to exist below a critical value of the inductance 
$\tilde{{\cal L}}_c$ where $5 \leq \tilde{\cal{L}}_c \leq 7$.
One has to choose, therefore, an $\tilde{{\cal L}}$ which is smaller than 
its critical value. 
To be sure that we are in the
CG phase we choose $\tilde{{\cal L}}=4$ and
$T=0.1$. 

The second harmonics have been obtained by employing
Monte Carlo simulations based on the standard Metropolis updating
technique.
While Monte Carlo simulations involve
no real dynamics, one can still expect that they give useful information on
the long-time behavior of the system.
We take $L=8$ and
$\omega=0.001$. The sample average is taken over 20-40 independent bond
realizations.
We set $h_{ac}=0.1$, corresponding to $\approx 0.016$ flux quantum per
plaquette. Smaller value of $h_{ac}$ turned out to leave the results
almost unchanged.

Fig. 2 shows the dependence of second harmonics
$|\chi_2|$, $|\chi_2|=\sqrt{(\chi'_2)^2 + (\chi''_2)^2}$,
on $h_{dc}$ in the FC regime.
Our calculations follow exactly
the experimental procedure of Heinzel {\em et al}\cite{Heinzel}.
First the system is cooled in the dc field $h_{dc}=1$
from $T=0.7$ down to $T=0.1$ which is below
the paramagnet-chiral glass transition temperature $T_{CG}\approx 0.17$
\cite{Kawa1}. The temperature step is chosen to be equal to 0.05. At each 
temperature, the system is evolved through
 2$\times 10^4$ Monte Carlo steps. When the lowest 
temperature is reached the dc field used in cooling is switched off and 
we apply the combined field given by Eq. (6).
Using Eq. (7) and (8) we monitor the second harmonics
(the technique for obtaing the ac susceptibility may be found in 
Ref. \cite{KawLi}) 
reducing the dc field from $h_{dc}=1$ to zero stepwise by
an amount of $\Delta h_{dc}=0.05$. $|\chi_2|$ reaches minimum at
the compensation field $h_{com}=0.7\pm 0.05$. At this point,
similar to the experimental findings\cite{Heinzel},
the intersection of $\chi'_2$ and $\chi''_2$ is observed. This fact 
indicates that at $H_{com}$ the system is really in the compensated state.
Furthermore, in accord with the experiments, at the compensation point the 
real and imaginary parts should change their sign\cite{Heinzel}.
Our results show that $\chi'_2$ changes its sign roughly at $h_{dc}=h_{com}$.
A similar behavior is also displayed by $\chi''_2$
but it is harder to observe due to a smaller amplitude of $\chi''_2$.

In conclusion we have shown that the finite-temperature CG phase  
is not spoiled by the screening.
The CE may be explained, at least 
qualitatively,
by using the CG picture of the ceramic superconductors.
The CE is shown to appear in the CG phase in which the PME is
present but not in the samples without the PME \cite{MSLi}.
\noindent

Financial support from the Polish agency KBN
(Grant number 2P03B-025-13) is acknowledged. 

\par

\end{document}